# Practical Evaluation of Low-Frequency Vibration Energy Harvesting for Creating Green RFID and IoT devices

Iván Froiz-Míguez*, Paula Fraga-Lamas*, Tiago M. Fernández-Caramés*

*Dept. of Computer Engineering
Centro de Investigación CITIC, Universidade da Coruña
A Coruña, Spain

*Abstract*—One of the main limitations for the development and deployment of many Green Radio Frequency Identification (RFID) and Internet of Things (IoT) systems is the access to energy sources. In this aspect batteries are the main option to be used in energy constrained scenarios, but their use is limited to certain cases, either because of the constraints imposed by a reduced-form factor, their limited lifespan, or the characteristics of the environment itself (e.g. operating temperature, risk of burning, need for fast response, sudden voltage variations). In this regard, supercapacitors present an interesting alternative for the previously mentioned type of environment, although, due to their short-term capacity, they must be combined with an alternative energy supply mechanism. Energy harvesting mechanisms, in conjunction with ultra-low-power electronics, supercapacitors and various methods to improve the efficiency of communications, have enabled the emergence of battery-less passive electronic devices such as sensors, actuators or transmitters. This paper presents a novel analysis of the performance of an energy harvesting system based on vibrations for Green RFID and IoT applications in the field of maritime transport. The results show that the proposed system allows for charging half of a 1.2 F supercapacitor in about 72 minutes, providing a stable current of around 210 µA and a power output of 0.38 mW.

*Index Terms*—energy harvesting, Green RFID, Green IoT, ultra-low power, battery-less sensor, supercapacitor, piezoelectric, harmonic motion

## I. Introduction

In recent years, the green digital evolution has contributed to improve energy efficient practices and automation with reduced environmental impact across multiple sectors thanks to the convergence of different technologies. One key technology of this transition is the Green Internet of Things (IoT) paradigm, in which elements of the physical world are interconnected with sensors or actuators following environmentally friendly practices.

The continuous improvement of semiconductor manufacturing technologies has led to enormous technological advances in small electronic devices, such as portable electronics, sensors and transmitters. Improvements have been made in terms of form factor and energy efficiency, thus enabling the reduction of the size of the devices by orders of magnitude.

This work has been funded by grant PID2020-118857RA-100 (ORBALLO) funded by MCIN/AEI/10.13039/501100011033.

However, there is still a significant barrier to energy storage, as the miniaturization of batteries is limited and other forms of energy storage (e.g., capacitors) are not as effective. In this respect, alternative sources of energy harvesting in conjunction with supercapacitors are helping in this miniaturization.

The generation of energy from ambient vibrations using piezoelectric energy harvesting devices has attracted research interest because of their ability to provide a constant energy source for Wireless Sensor Networks (WSNs) [1]. In this regard, harmonic motion sources that produce constant frequencies are widely used in many sectors of society and represent an efficient way of obtaining energy through energy harvesting modules [2].

## II. Harvesting Energy from Vibrations

### A. Basics on Energy Harvesting

Energy harvesting mechanisms allow for obtaining energy from external environmental sources, usually providing a very small amount of energy. The main characteristics present in energy harvesting applications are low data rate, low duty cycle and ultra-low power. Harvesting mechanisms can be found in Radio Frequency Identification (RFID) and WSN-based applications [3]. For example, in passive RFID tags [4], wearable devices, flexible electronics or smart buildings [5].

In certain situations, the conventional storage of energy by batteries is a constraint. One of the main concerns with batteries is degradation, which significantly reduces their lifespan. Moreover, specially for WSNs, batteries are hard to access for recharging or replacement. Furthermore, certain portable devices have a reduced form factor that does not allow for incorporating batteries [6].

In this regard, capacitors have longer lifespan, deliver charge faster and have no memory effect, but they store less charge. In addition, the enormous reduction in the size and power consumption of circuits has led to a research effort focused on on-board power sources that can replace batteries [7].

Supercapacitors have a significantly higher capacity than regular capacitors while retaining the same reduced size and operating voltage, thus bridging the gap between electrolytic capacitors and rechargeable batteries. Moreover, supercapacitors allow for storing 10 to 100 times more energy per unit of volume or mass and can accept and deliver charge much faster than batteries.

While batteries are used for long-term compact energy storage, it is possible to overcome this limitation of capacity in supercapacitors by incorporating energy harvesting circuits that can extract the maximum amount of energy from an external source via an energy transducer.

*B. Types of Energy Harvesting Transducers*

Alternative energy sources are available in the surroundings of most deployment sites. Examples of such energy sources include mechanical (e.g., vibrations, deformations), thermal (e.g., temperature gradients or variations), radiant (e.g., solar, infrared or Radio Frequency (RF) energy), and chemical energy (e.g., chemistry, biochemistry) sources, among others. These external energy sources are converted to electricity through different kinds of transducers, and the obtained electricity can power devices or can be stored for further use. Some of the common types of harvesting sources and transducers are photovoltaic, thermoelectric, piezoelectric, electromagnetic and electrostatic.

*C. Vibration-Based Energy Harvesting System*

The solutions proposed in [2] and [5], despite providing a theoretical and practical analysis of vibration-based harvesting modules in different environments, do not provide an exhaustive analysis of the charge capacity that can be achieved with this type of harvesting. Such an analysis is of particular interest in cases where the output power of the harvester is not enough to directly power the device (i.e., the power employed by RF modules in wireless transmissions may have a much higher consumption compared to the rest of the electronics).

The objective of the research presented in this paper is to analyze the feasibility of creating an energy harvesting system based on piezoelectric components that generate energy from vibrations to power different types of low-power devices/sensors. Such a system should capture ambient mechanical energy and convert it into electrical energy through the oscillation of a laminated piezoelectric material, to finally charge a supercapacitor.

Vibration sources can be generated through several external agents such as ocean waves, bridges, industrial equipment, human motion or automobiles, among others. Traditionally, vibration energy is dissipated into heat waste by the damping elements of the systems. Piezoelectric transducers, transform this mechanical energy through an internal electric polarization that is produced by the exposure to mechanical strain caused by vibrations. Figure 1 shows the operation of a vibration harvester and how the motion energy is rectified in electricity (a) as well as the different elements that compose this type of device (b).

*D. Definition of the Evaluation Scenario*

When designing an energy harvesting system it is important to consider that the efficiency is directly related to the resonance frequency, so to obtain the best results, the operating excitation frequency should be within the range of the resonance frequency of the harvester.

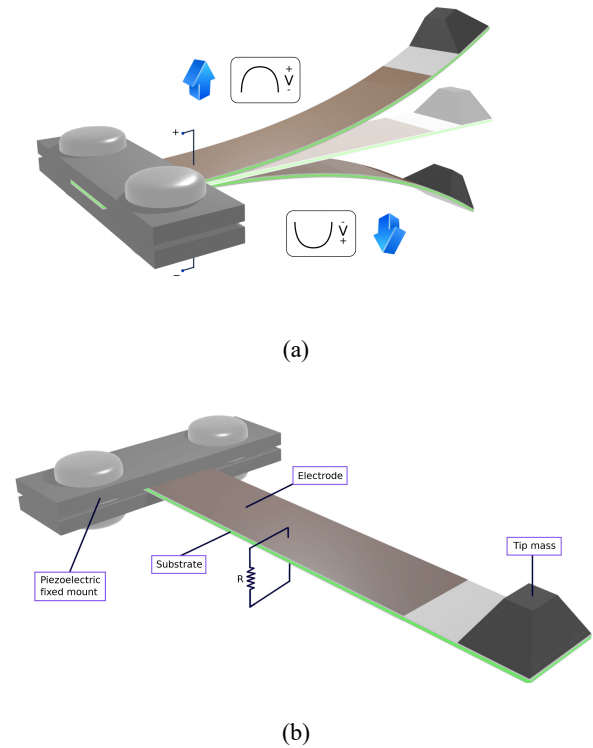

Fig. 1: Basic operation of an energy harvesting vibration module (a) and schematic diagram of a piezoelectric energy harvester (b).

However, resonance inherently induces a narrow bandwidth, which means that a slight shift of the resonance frequency will cause a drastic drop in power output. Therefore, vibration-based piezoelectric energy harvesting systems operate in a short bandwidth range close to the resonant frequency. Moreover, resonance restricts the miniaturization of vibration harvesters due to the size effect (i.e., smaller sizes lead to higher resonance frequencies).

Another key factor related to resonance is the environment in which the harvester operates. While environmental motion sources are characterized by a wide range of frequencies, mechanical vibration sources can have a much smaller range, like those related to rotational motion (i.e., generated by shafts, wheels or gears), especially if they present a constant angular velocity and acceleration. In this regard, rotary energy is widely used in multiple sectors. One of such sectors is transportation, since most vehicles use engines or other propulsion systems that generate rotational kinetic energy. There are certain engines that operate at constant speeds most of the time, (i.e., when using cruise control), which is widely used in airplanes or ships for a more efficient use of fuel. This type of environment is optimal for high performance harvesting due to the short bandwidth it presents.

In the specific case of maritime transportation, ships are subjected to considerable vibrations during operation, either by the excitation of the propeller or by external factors. Such vibrations directly affect both the machinery and the habitability of the environments. In order to avoid vibration-related issues, different standards have been developed like ISO 20283-5:2016, ISO 20816-1:2016 or MIL-STD-167-1A. It is possible to use such standards to simulate and evaluate the performance achieved by vibration harvesting modules when they are under the circumstances defined by each standard.

In particular, MIL-STD-167-1A [8] specifies procedures and establishes requirements for environmental and internally excited vibration testing of naval shipboard equipment installed on ships with conventionally shafted propulsion. Such a standard proposes a set of stress tests on the equipment to measure the response to environmental vibrations caused by hydrodynamic forces on the propeller blades, which interact with the hull and with other external sources. Table I indicates the signal amplitude for different frequencies and the displacement that the equipment must be able to withstand according to MIL-STD-167-1A.

TABLE I: Vibratory displacement of environmental vibration.

| Frequency range (Hz) | Single amplitude (mm) |
|---|---|
| 4 to 15 | 0.762 |
| 16 to 25 | 0.508 |
| 26 to 33 | 0.254 |

An Inertial Measurement Unit (IMU) can be used to measure the displacement produced by vibrations on the surface based on its acceleration in the Z-axis. To simulate the displacements allowed within the frequency ranges indicated in Table I, a vibration generator is used to provide a periodic harmonic motion at a specific frequency. Considering the harmonic motion of a periodic waveform, there is a mathematical relationship between frequency, displacement, velocity and acceleration:

$$D = \frac{A}{(2\pi F)^2} \quad (1)$$

Thus, Equation 1 defines the calculation of the displacement $D$, being $A$ the acceleration (usually expressed in $m/s^2$) and $F$ the vibration frequency. An IMU is used to provide acceleration in g-units, therefore it is necessary a conversion to be able to obtain displacement distances (1 $g$ = 9.80665 $m/s^2$). Moreover, assuming that the frequency and acceleration stay constant due to harmonic motion, it is possible to express the different variables in amplitude (zero-to-peak) or maximum value (peak-to-peak), as it is illustrated in Figure 2.

### III. SYSTEM DESIGN AND EXPERIMENTAL SETUP

#### A. Overview of the system

There are over 200 piezoelectric materials that could be used for energy harvesting, with the appropriate ones being selected for each application depending on the type of mechanical stress applied and the characteristics of the environment. In this regard, the ceramic lead zirconate titanate, also known as PZT, is still the most commonly used material for piezoelectric harvesting. The popularity of PZT is due to the fact that it is one of the most efficient and cost-acceptable materials.

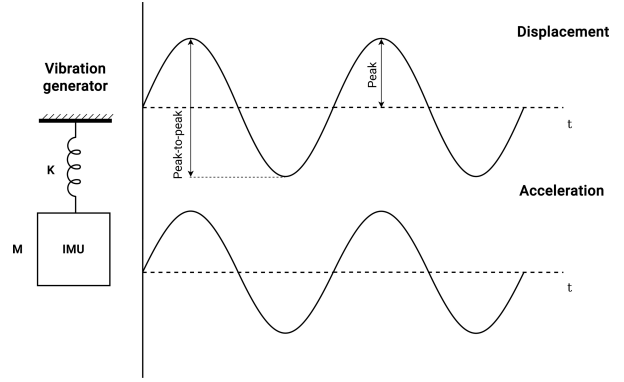

Fig. 2: Simple harmonic motion, where it is possible to define an instantaneous displacement or acceleration at a given time.

One characteristic of the piezoelectric modules is that they generate an AC current, which requires the use of a rectifier in order to be used directly on low-power hardware. Such rectifiers must be highly efficient in order to minimize energy loss, since vibration harvesters generate very low currents (in the order of µA).

Considering the previous observations, Figure 3 shows the circuit designed for the proposed development. As it can be observed, the energy harvesting module is connected directly to a signal rectifier which transforms the signal to DC and provides a stable regulated voltage at the output. Such an output is then used to directly charge a supercapacitor.

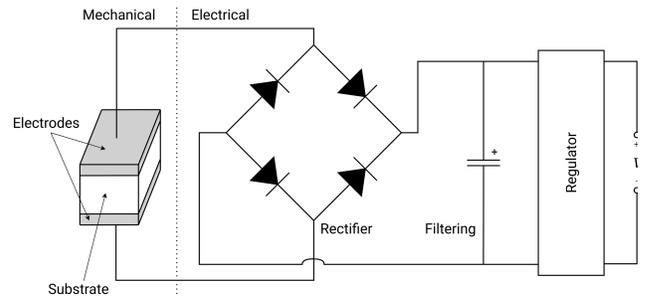

Fig. 3: Simplified electrical circuit for the designed system.

#### B. Selected Harvester and Measurement Tools

The piezoelectric module selected for the proposed development is a PZT manufactured by Midé. The selection was performed by considering the frequency ranges indicated in Table I. Among the piezoelectrics distributed by the manufacturer, only 5 models are designed to operate at such low frequencies, being the model PPA-2011 the one with the best performance.

The following are the main characteristics of the components used by the devised system, including the tools used for measuring its performance (Fig. 4 shows a picture of such tools):

- Piezoelectric harvester: Midé PPA-2011 vibration energy harvester with 190 nF of capacitance, voltage range of ± 120 V, resonance frequency of 178 Hz and a size of 71x25.4 mm.
- Rectifier module: DC1459B-A based on LTC3588.
- Supercapacitor: voltage 2.7 V, capacitance 1.2 F.
- Digital oscilloscope: Hanmatek DOS1102 with 110 MHz of bandwidth and a sampling rate up to 1 GS s$^{-1}$ It is used for analyzing the AC signal produced by the piezoelectric module and by the vibration generator.
- Joulescope: precision DC energy meter. This tool has a resolution of 1.5 nA and a sampling rate of 2 MS s$^{-1}$. It is used as a data logger of the supercapacitor charge.
- Current ranger: high precision current meter that allows for capturing fast current transients with a resolution ranging from mA to nA up to 3.3 A. It is used to determine the value of the stable current produced by the harvester.
- Signal generator: JDS6600. Dual-channel signal generator with a frequency range between 1 Hz and 100 MHz, and a signal amplitude between 2 and 20 Vpp, this device generates electrical signals with predefined properties such as amplitude, frequency and wave shape. It provides a load capability of 150 mA with 50 Ω of impedance. It is used to simulate an external source of vibrations.
- Vibration generator: 3B Scientific 1000701. It is used to generate mechanical waves to study oscillations and resonance. It provides a maximum amplitude of 5 mm, 8 Ω of impedance and up to 20 kHz. It is the external motion source used for the harvester.
- IMU: MPU6886. It is a 6-axis sensor built-in on the m5-stack board.

## IV. Experiments

### A. Experimental setup and evaluated scenarios

The experiments were carried out in two different stages. The first one consisted in analyzing the response of the harvesting module under different oscillation signals, while the second one consisted in analyzing the time of charge of the supercapacitor through the rectifier module, as well as measuring the current provided by the piezoelectric on a regular operation.

In order to analyze the performance of the harvester under different types of vibrations, the signal generator was used to supply a signal of a given amplitude and frequency to the vibration generator. The energy harvesting module was attached to the vibration generator with neodymium magnets, and the output signal provided by the piezoelectric module was analyzed through the oscilloscope.

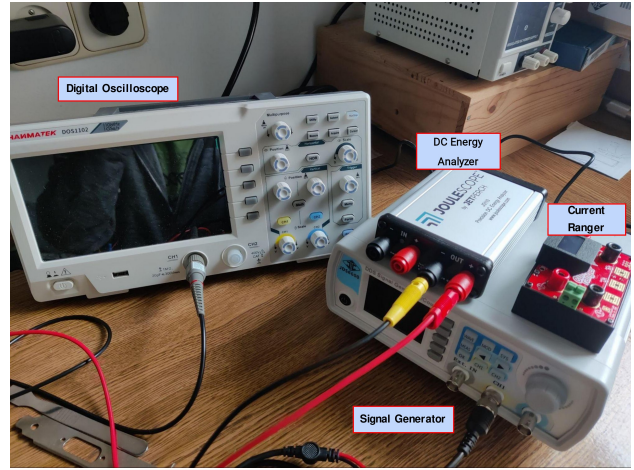

Fig. 4: Instrumentation involved in the performed measurements and for signal generation.

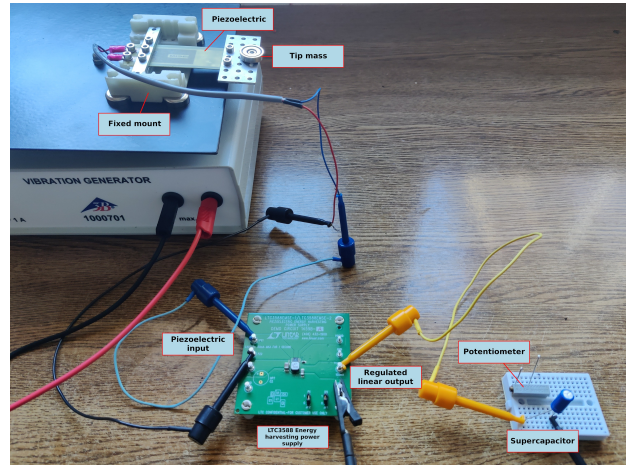

Fig. 5: Hardware setup used during the experiments (the rectified and regulated output of the power supply is directly connected to the supercapacitor or to the potentiometer).

Although it is not a mandatory component, it is highly recommended to use a clamper kit with the energy harvesting module as vibration harvesters are very sensitive to variations in position. The used harvesters include a fixed part (at the base of the piezoelectric) that allows the module to oscillate freely. The harvester also offers the possibility of adding weight at the tip to increase the oscillation and generate more energy. Figure 5 shows the piezoelectric module mounted in the clamper kit attached to the vibration generator.

After setting up the harvester and the measurement tools, it is possible to analyze the signal at which vibration generator operates. Note that, although the signal is supplied by the signal generator, the vibration generator has a low impedance (8 Ω) and the signal generator is not capable of supplying power to work with such a low impedance, so there is a notable drop in the used voltage. Moreover, it is important to measure the displacement generated on the vibration generator through an IMU and check that the values are within the ranges established in Table I.

Two different scenarios were defined: "A" and "B". In scenario A a signal of 23.5 Hz and 10 Voltage peak-to-peak (Vpp) was supplied, while 23.5 Hz and 16 Vpp were used for the scenario B. Table II shows the acceleration and displacement values that occur in both scenarios. Moreover, the voltage obtained at the output of the vibration generator is indicated in Table III.

TABLE II: IMU acceleration and displacement obtained in both scenarios.

| Scenario | A | B |
|---|---|---|
| Gpp (g) | 0.52 | 0.98 |
| Dpp (mm) | 0.210 | 0.405 |
| Frequency (Hz) | 23.5 | 23.5 |

TABLE III: Voltage peak-to-peak obtained at the output of the vibration generator and piezoelectric for both scenarios.

| Scenario | A | | B | |
|---|---|---|---|---|
| | Generator | Piezoelectric | Generator | Piezoelectric |
| Vpp (V) | 1.367 | 22.66 | 1.797 | 26.56 |
| Frequency (Hz) | 23.5 | 23.5 | 23.5 | 23.5 |

### B. Determining the Optimum Performance

To evaluate the performance of the harvesting module, it is first essential to assemble correctly the clamper kit. In addition, adding tip mass to the harvester is crucial for optimization, since it will have a direct impact on the resonance and oscillation frequency of the module and therefore on the maximum generated energy. Fig. 5 shows the assembly of the tip weight, which includes the following components:

- A neodymium magnet (5.7 g).
- A connection plate (8 g).
- Nuts (3 x 0.4 g).
- Screws (3 x 0.7 g).

With a total weight of 17 g and the resonance frequency set at 23.5 Hz, the output of the harvesting module for the inputs defined by scenarios A and B, are represented in Figures 6 and 7, respectively. As it can be observed in such Figures, the voltages obtained with the harvester are relatively high, considering the reduced displacement provided by the vibration generator in both scenarios. Apart from determining the optimum resonance frequency, the tip mass and the correct assembly in the clamper kit are crucial to obtain the best performance.

### C. Results and Performance

Now that the optimum mounting configuration and resonance frequency of the harvester module have been determined, it is possible to analyze the performance of the harvesting circuit. For such a purpose, the output of the harvester module is connected to the voltage rectifier/regulator kit.

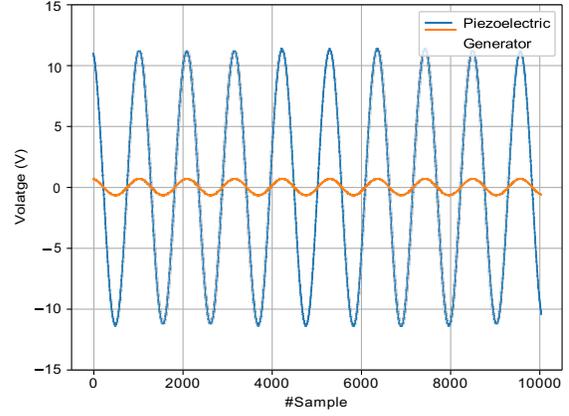

Fig. 6: Signal obtained in the vibration generator and piezoelectric for scenario A.

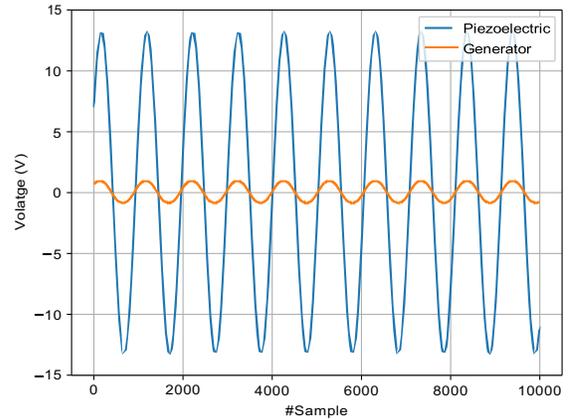

Fig. 7: Signal obtained in the vibration generator and piezoelectric for scenario B.

Initially, a variable load of maximum 20 kΩ (a potentiometer) was used to analyze the generated current. After determining the delivered current offered by the harvester module, it was measured the time spent to charge a supercapacitor of 1.2 F (all the mentioned components are shown at the bottom of Fig. 5).

As it can be observed in Figures 6 and 7, the voltage generated at the output of the piezoelectric module is relatively high (22.66 and 26.65 Vpp for scenarios A and B, respectively). The LTC3588 waveform rectifier/regulator module integrated in the DC1459B-A is optimized for high output impedance energy sources with an input protective shunt for input voltages greater than 20 V. At the output of the DC1459B-A module, a voltage of 1.8 V was selected for direct use on the supercapacitor.

Current consumption was analyzed with the current ranger. Thanks to the variable resistance provided by the potentiometer, it was possible to analyze the current at which the DC1459B-A module provides a stable output. In scenario A it remained stable at 10.7 kΩ with values between 165 and 169 µA, while in scenario "B" it remained stable at 8.5 kΩ with values ranging between 209 and 213 µA.

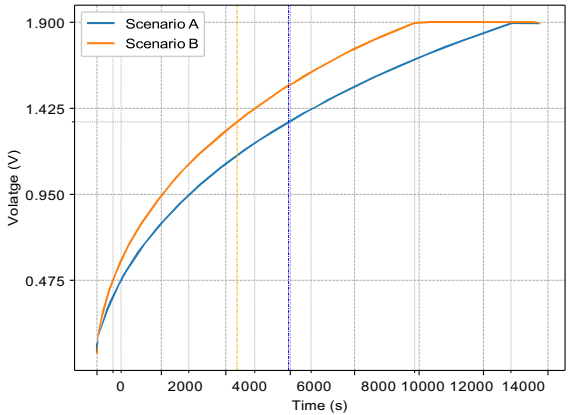

Fig. 8: Supercapacitor charge curve obtained in the scenario A and B, vertical asymptotes represent a charge of 50% of the supercapacitor for each scenario.

Regarding the charge of the supercapacitor, Figure 8 shows the evolution over time in both scenarios, it is important to note that although the output selected for the rectifier/regulator module was 1.8 V, both curves show that the output becomes flat at approximately 1.9 V. According to the datasheet, the output voltage for the 1.8 VOUT mode is in the range of 1.71 to 1.89 V. Moreover, reaching such a voltage level takes slightly more than 210 minutes for scenario A and slightly more than 160 minutes for scenario B.

### D. Considerations and Key Findings

Vibration energy harvesting is currently limited to applications where the vibration environment maintains a relatively constant frequency. In practice, most vibration environments exhibit variability and consist of multiple frequency components. There is also a number of applications that can benefit from harnessing kinetic energy from shock events and other transient events. Nonetheless, traditional piezoelectric energy harvesting methods are not efficient for harnessing this type of energy.

An alternative to overcome the limitation of the wide range of frequencies that are present in nature is to create an array of vibration harvesters, each tuned to a particular frequency in the covered range [9]. However, the cost of such an array of piezoelectric harvesters is still high. Moreover, another consideration to keep in mind is the high output impedance, which produces relatively high output voltages at low electrical current which may not be adequate for certain applications.

Regarding the supercapacitors, as it was shown in Figure 8, although the load of the supercapacitors is logarithmic and it takes time to charge it to high values, its loading is much faster at the beginning and can be charged to 50% of its capacity in approximately 100 and 72 minutes in scenarios A and B, respectively. Moreover, unlike batteries, capacitors have no memory, which means that these charging times will be maintained almost constant throughout their entire life span.

### E. Conclusions

This paper presented an energy harvesting system that rectifies vibrations using piezoelectric modules and is focused on rotary movements at low frequencies. Specifically, the devised system has been designed for ship environments, where large ships are complex structures with multiple factors to monitor and control, which make them really interesting for Green RFID and IoT applications. After presenting the design of the system, two scenarios were defined, which presented different amplitude values, both within the margins established by the MIL-STD-167-1A standard. In such scenarios the performed experiments measured powers of approximately between 0.3 and 0.4 mW and showed that it is possible to charge a 1.3 F supercapacitor by a half in 72 minutes in the best case.

Therefore, the devised system can be used as a battery replacement for many applications that make use of low-power devices, thus increasing the number of environments in which it can be deployed as well as its lifetime while maintaining a small form factor. However, it is necessary to guarantee a minimum vibration time close to the harvester resonance to obtain better performance, which is feasible in this type of rotating motion environments.